\documentclass[preprint2]{aastex}
\usepackage{graphicx}
\usepackage{txfonts}
\usepackage{natbib}
\newcommand{\EWmin}{EW$_{\rm min}$}
\shorttitle{ESO VLT optical spectroscopy of BL Lac objects.}
\shortauthors{Sbarufatti et al.}
\begin{document}
\title{ESO VLT optical spectroscopy of BL Lacertae objects. III. An extension of the sample.}
\author{B. Sbarufatti}
       \affil{INAF, Istituto di Astrofisica Spaziale e Fisica Cosmica di Palermo}
       \affil{Via Ugo La Malfa 153, I-90146 Palermo, Italy}
       \email{sbarufatti@ifc.inaf.it}
\author{S. Ciprini}
       \affil{
         1. Physics Department University of Perugia, and I.N.F.N.
         Perugia Section,\\
         Via A. Pascoli, I-06123 Perugia, Italy\\
         2. Tuorla Observatory, University of Turku,\\
         V\"ais\"al\"antie 20, FIN-21500 Piikki\"o, Finland} 
\author{J. Kotilainen}
       \affil{Tuorla Observatory, University of Turku}
       \affil{V\"ais\"al\"antie 20, FIN-21500 Piikki\"o, Finland}
\author{R. Decarli, A. Treves, A. Veronesi}
       \affil{Universit\`a dell'Insubria}
       \affil{Via Valleggio 11, I-22100 Como, Italy}
      \and
\author{R. Falomo}
       \affil{INAF, Osservatorio Astronomico di Padova}
       \affil{Vicolo dell'Osservatorio 5, I-35122 Padova, Italy}
\begin{abstract}
We present results of an ongoing program at the ESO VLT for spectroscopy 
of BL Lac objects lacking a firm redshift estimate
and here we report on 15 objects. For 11 sources we confirm 
the BL Lac classification, and we determine new redshifts for 3 objects, 1 
with weak emission lines (PKS 1057$-$79, $z=0.569$) and 2 with absorptions 
from the host galaxy (RBS 1752, $z=0.449$; RBS 1915, $z=0.243$); moreover 
a sub Damped Lyman Alpha (sub-DLA) system 
is detected in the direction of the BL Lac PKS 0823$-$223 ($z\geq0.911$). For 
the remaining 8 BL Lacs, from the very absence of absorption lines of the host 
galaxy, lower limits to the redshift are deduced with $z_{\rm min}$ in the 
interval $0.20-0.80$.  
The remaining three sources are reclassified as a FSRQ (PKS 1145$-$676, 
$z=0.210$; TXS 2346+052, $z=0.419$) and a misclassified galactic star 
(PMNJ 1323$-$3652).
\end{abstract}


\section{Introduction}
Blazars dominate the scene of extragalactic gamma ray astronomy as space borne 
missions and \v{C}erenkov atmospheric telescopes have shown. 
BL Lac objects ( BL Lacs or BLLs)
are a main sub-class of \textit{blazars} that by definition exhibit featureless
spectra or very weak lines, most probably because of the relativistic 
enhancement of the continuum.
Surely, the recent launch of AGILE and GLAST  gamma-ray observatories 
and the upgrading of the existing ground-based \v{C}erenkov telescopes will 
significantly increase the interest in the BL Lac objects.
The determination of the redshift is mandatory in order to characterize these 
sources (e.g. \ to determine nuclear and host galaxy luminosity of the 
sources), but it is arduous, in most of the objects, to detect the weak 
spectral features over the continuum. Indeed,  a redshift determination exists 
for only about half of the known BLLs. The detection of the weak spectral 
lines necessarily requires the use of $8-10$ meters telescopes. 
In addition to the issue of redshift, high S/N and spectral resolution
are also of importance for detecting the host galaxy independently of 
imaging, since its emission may appear superposed to the non-thermal 
continuum of the nucleus.
Absorption lines may be related to the host galaxy itself, to the 
intergalactic medium and to the interstellar medium of our galaxy and its 
halo. The emission lines are the most direct probe to the physical conditions 
around the nucleus.

We have an ongoing program for BL Lac's spectroscopy at the European Southern 
Observatory (ESO) $8-$ meter Very Large Telescope (VLT), 
which utilizes the telescope in service mode under non-optimal seeing 
conditions.
The results of the first three runs (2003 and 2004) referred to 35 BLLs. 
We measured new redshifts for 17 sources, while for the rest of the 
objects we have given upper limits using a technique specifically designed 
for this project. Details are given in \citet[][ S05 and S06 in the 
following]{sbarufatti05_vlt1, sbarufatti06_vlt2}, together with the criteria 
on the sample selection.

In this paper we present the spectra of 15 objects observed in 2006 
(Guest Observer run: ESO 077.B-0045).
Data reduction and analysis procedures are described in section 2. Results are 
reported in section 3 along with specific comments about each source. Summary 
and conclusion are given in section 4. Throughout this paper we assume the 
following cosmological parameters: H$_0$=70 km s$^{-1}$ Mpc$^{-1}$, 
$\Omega_\Lambda$=0.7, $\Omega_{\rm m}$=0.3.

\section{Observations and data analysis}
 
The observations (Table \ref{tab:journal})
were performed between 2006 March through August in Service Mode at 
the ESO VLT UT2 (Kueyen) telescope, equipped with the FOcal Reducer and low 
dispersion Spectrograph (FORS1), using the 300V+I grism combined with a 
2\arcsec{ } slit, yielding a dispersion of 112 \AA~mm$^{-1}$ (corresponding to 
2.64 \AA~pixel$^{-1}$) and a spectral resolution of 15  \AA~covering the 
3800--8000 \AA~range. 
The seeing during observations was in the range 0.5-2.5\arcsec, with an 
average of 1\arcsec.

We performed data reduction using IRAF\footnote{IRAF (Image Reduction and 
Analysis Facility) is distributed by the National Optical Astronomy 
Observatories, which are operated by the Association of Universities for 
Research in Astronomy, Inc., under cooperative agreement with the National 
Science Foundation.} \citep{tody86,tody93}, following the standard procedures 
for spectral analysis. This includes bias subtraction, flat fielding, and  
removal of bad pixels. For each target, we obtained three spectra for an 
optimal correction of the cosmic rays and to check for the reality of 
weak spectral features. The individual frames were then combined into 
a single average image. Wavelength calibration was performed using the spectra 
of a He/Ne/Ar lamp, 
resulting in an accuracy of $\sim$3 \AA \ (rms). From these calibrated final 
images, we extracted the one-dimensional spectra inside a 
2\arcsec$\times$6\arcsec aperture, adopting an optimal extraction algorithm 
\citep{horne86} to improve the Signal to Noise ratio (S/N). 

As a part of a fill-in program, our observations did not require optimal 
photometric conditions. However, the sky was clear for most of the 
observations. This enabled us to perform a spectrophotometric calibration of 
the data using standard stars \citep{oke90}. We estimate an uncertainty
of the order of 10\% in the flux calibration because of the not optimal sky 
condition. Flux losses due to the slit not being oriented along the parallactic 
angle are negligible with respect to the flux calibration uncertainties.
All the spectra were dereddened following the extinction law by 
\citet{cardelli89} and assuming the E$_{\textrm{B-V}}$ values computed by 
\citet{schlegel98}. 
 
\section{Results}\label{sec:results}

In Figure \ref{fig:spectra} we give the optical spectrum for each source. In 
order to make apparent the shape of the continuum and the faint spectral 
features, we report both the flux calibrated and the normalized spectra.
Intrinsic and intervening spectral features are identified
with atomic species. Absorptions caused by the Galactic 
interstellar medium are indicated with ISM for simple atomic species, and with 
DIB (Diffuse Interstellar Band) for complex molecules. The Earth symbol is used
to mark telluric absorptions. All spectra can be retrieved in electronic form 
at \texttt{http://www.oapd.inaf.it/zbllac/}, where all the results of our program are 
archived.

\subsection{Continuum emission and host galaxy contribution}\label{sec:cont}

The optical spectrum of a BL Lac can be
described, in a first approximation, as the superposition of two components. 
The first one is a non-thermal continuum emitted by the active nucleus. 
The second is the contribution from the stars and the ISM of the
BLL host galaxy. Extensive studies in the past \citep[e.g.][]{urry00} 
have shown that these galaxies are usually giant ellipticals 
the optical magnitudes of which follow a narrow Gaussian distribution centered 
around M$_{\rm R}^{\rm host}=-22.9\pm0.5$ \citep{sbarufatti05_hst}. In most 
cases the host galaxy signature was not detected in our spectrum because it 
was too faint with respect to the nuclear component. For these objects, we 
performed a fit of the optical continuum with a simple power-law model 
($F_\lambda \propto \lambda^{-\alpha}$). In two cases however (RBS 1752 and RBS 
1915) the host galaxy spectral features were detected, and we performed a fit 
to the spectrum using a two component model (see Fig. \ref{fig:specdec}): a 
power-law plus an elliptical galaxy spectrum as in the template given in 
\citet[][see also S06 for details on the fitting procedure]{kinney96}.
The results from our fits are reported in Table \ref{tab:fits}, where for each 
object we give the power-law index, the object class \citep[High 
Energy Peaked BLL, HBL; Low Energy Peaked BLL, LBL, see][for a definition]
{padovani95b} the apparent magnitude, and the extinction 
coefficient. For RBS 1752 and RBS 1915 we report also the absolute magnitudes 
of the host galaxies corrected for the aperture effect which were 
M$_{\rm R}= -23.3$ and M$_{\rm R}$= -22.4 respectively, in agreement with the 
expected distribution.

\subsection{Line detection and redshift determination}

Since emission and absorption lines in a BLL spectrum can be very faint, their 
detection can be a difficult task. Using the same technique presented in S06, 
we estimated the minimum detectable equivalent width \EWmin~ for each spectrum,
and considered all features with equivalent width (EW) above this threshold as 
line candidates which were then carefully inspected for identification or 
rejection. 
\EWmin ~ values for each spectrum are given in table 1, while line 
identifications, Full Width Half Maximum (FWHM) and EW are reported in table 
\ref{tab:lines}. The continuum and line properties confirmed the BL Lac 
classification for 12 objects. In section \ref{sec:notes} we report notes on 
the individual sources.

\subsubsection{Redshift lower limits}
Most of the confirmed BLL in our sample show featureless spectra, despite the 
high S/N reached using VLT. As extensively discussed in S06 
(4.2.4), it is possible to estimate a lower limit to the redshift of such 
sources, knowing the \EWmin ~ and the nucleus apparent magnitude and exploiting
the assumption that BL Lac hosts can be considered as candles 
\citep{sbarufatti05_hst}. 
We remark here that Equation 1 in S06 contains a typographical error, which was
noted also by \citet{finke08}. The correct expression is:
\begin{equation}\label{eq:1}
\mathrm{EW}_{\rm obs}=\frac{(1+z) \times \mathrm{EW}_0}{1+\rho / A(z)}
\end{equation}
where EW$_{\rm obs}$ is the observed equivalent width, EW$_{0}$ is the 
equivalent width of the feature in the host galaxy template \citep{kinney96}, 
$\rho$ is the nucleus-to-host flux ratio, $z$ is the redshift and $A(z)$ is 
the aperture correction.

We applied this procedure to all featureless spectra in the sample, obtaining 
the lower limits reported in section \ref{sec:notes}.

\subsection{Notes on individual objects.}\label{sec:notes}

\paragraph{PKS 0019+058}   
This radio selected source \citep{condon74} was first classified as a BLL
by \citet{fricke83}, based on a featureless optical spectrum. We observed this 
object in two epochs separated by about a month, noticing an optical 
variability of 0.7 mag ($R$ band) and an evolution of the spectral index from 
0.65 to 0.76. The reality of the optical variability is fully confirmed by 
the $R$-band images exposed prior to the spectra which enable direct 
photometry. The optical variability was known, both in magnitude 
\citep[$V= 19.2$ and $V>21$ in][ respectively]{fricke83, abraham91} 
and in spectral index \citep[$\alpha_{\nu}$=0.8 and 0.94 in ][]{fricke83,
chen05}. We find no intrinsic feature with EW $>$ \EWmin~ in any of the two 
spectra. The NaI$\lambda$5891 absorption from the Galaxy ISM is detected in 
both spectra with EW $=0.43$ \AA~ (July 12) and 0.66 \AA~ (August 8).
A marginally significant excess around 5600 \AA~ is present in both 
observations, but it is most probably spurious because of a residuals left  
after the subtraction of a nearby atmospheric emission  line.  
The most stringent redshift lower limits, obtained from the August 2006 
spectrum is $z>$0.64.
     
\paragraph{GC 0109+224}	      

This source was discovered in radio observations by \citet{davis71}, and 
subsequently classified as a BL Lac by \citet{owen77}. Strong flux and 
polarization variability was reported both in the radio and optical band 
\citep[e.g.][]{katajainen00, ciprini03, ciprini04}. Optical imaging by 
\citet{falomo96} and \citet{nilsson03} failed to detect the host 
galaxy, while the detection reported by \citet{wright98} is dubious,
since it has not been confirmed by any subsequent observation. 
The lower limit on the redshift proposed by \citet{falomo96} is $z>$ 0.4. 
Previous optical spectroscopy by \citet{wills79,falomo94,sbarufatti06_eso36}
performed with 2--4 m class telescopes showed featureless spectra,
while \citet{healey08} report z=0.265 for this object based on an 
unpublished optical spectrum.
The lower limit to the redshift as deduced by \citet{sbarufatti06_eso36} 
was based on an \EWmin$= 0.43$ \AA~ that yielded  $z > 0.18$. The considerably 
higher S/N spectrum we obtained with VLT has \EWmin$=0.09$ \AA, which in turn 
implies $z>$ 0.25, which improves the previous spectroscopical limit, but 
it is still less stringent than the imaging limit.

\paragraph{RBS 0231} 	      

This X-ray-selected object \citep{voges99} was classified as a BLL by 
\citet{schwope00}, and as a HBL by \citet{brinkmann00}. No previous
optical spectroscopy was published. The VLT spectrum is featureless, 
with \EWmin$=1.27$ \AA, which implies $z>0.41$.

\paragraph{PKS 0823$-$223}	      

This radio selected BL Lac \citep{allen82} is characterized by a number of 
absorption lines in the UV and optical band \citep[e.g.][]{rao06, 
meiring07, falomo90, veron90, falomo94}. These features are consistent with 
the presence of a sub-Damped Lyman $\alpha$ Absorber at $z = 0.91$. In our VLT 
spectrum we detect the absorption features of 
FeII $\lambda\lambda$2373.7, 2383.2, 2585.9, 2599.4, MgII 
$\lambda$2798 and MgI $\lambda$2852 at the same redshift. 

\paragraph{PKS 1057$-$79}	      

This radio selected object \citep{shimmins81} was proposed as a
counterpart for the $\gamma$-ray source 2EGS 1050$-$7650 by 
\citet{tornikoski02}. No previous optical spectroscopy was published. 
Our VLT spectrum shows several emission lines ([OIII] 
$\lambda\lambda$4959,5007, [NeIII] $\lambda$3868, MgII $\lambda$2798), at 
$z=0.581$. Since the FWHM of the MgII line (see table \ref{tab:lines}) 
is in excess of $1000~km~s^{-1}$ we propose to classify this object as a 
broad line AGN. 

\paragraph{PKS 1145$-$676}	      

This radio source was classified as a quasar due to its point-like appearance 
by \citet{white87}. The flat radio spectrum and optical variability 
\citep{beasley97, costa02} prompted a \textit{blazar} classification. We 
detect several emission lines ([OII] $\lambda$3727, H$\beta~ \lambda$4861, 
[OIII] $\lambda\lambda$4959,5007, H$\alpha~ \lambda$6563, [NII] $\lambda$6585) 
at $z= 0.210$. The observed EW are in the range 4--25 \AA, pointing towards a 
FSRQ classification. The FWHM of the  H$\beta~$ line could indicate 
that this source is a narrow lines object, but the fact that we are unable 
to measure the H$\alpha~$ FWHM due to the blending with the [NII] lines 
permits only a type $\sim1.9$ classification.

\paragraph{OM 280}	

This radio selected source \citep{colla72} was classified as a BLL due 
to its featureless spectrum by \citet{strittmatter74}. Subsequent optical 
spectroscopy by \citet{rector01} also failed to discover any intrinsic 
spectral feature. The host galaxy was not detected in deep HST imaging 
\citep{urry00}, implying $z>0.63$ \citep{sbarufatti05_hst}. The VLT 
spectrum is featureless, with \EWmin$=0.35$ \AA, which implies a redshift 
lower limit of 0.20, less stringent than the one from imaging.	      

\paragraph{PMN J1323$-$3652}       

This radio selected source was classified as a candidate BL Lac by the WGA 
catalog \citep{wga2000} and the Deep X-ray Radio Blazar survey 
\citep[DXRBS,][]{landt01}, that also provided a featureless 
optical spectrum. However, our VLT optical spectrum clearly shows the 
absorption lines and the characteristic shape of a Galactic F-type star. This 
indicates a mis-identification of the optical counterpart of this source.

\paragraph{OQ 012}	

This radio selected BLL \citep{weiler80} showed a featureless optical spectrum
in observations by \citet{falomo94}. \citet{richards04} gave a photometric 
redshift estimate  $z=0.475$ based on data from the Sloan 
Digital Sky Survey (SDSS). Our VLT spectrum shows an absorption line by 
Galactic ISM (NaI $\lambda$5891), but no intrinsic features. The \EWmin~ is 
0.31 \AA, which implies $z>0.65$, inconsistent with the photometric estimate.

\paragraph{PMNJ 1539$-$0658}       

This radio source \citep{griffith95} was classified as a BL Lac in the DXRBS 
\citep{landt01}, which also provided a featureless optical spectrum. In the 
VLT spectrum we detect the NaI $\lambda$5891 absorption line from the Galaxy, 
but no intrinsic features. The minimum detectable EW is 0.61 \AA, which 
implies $z>0.80$.

\paragraph{PKS 1830-589}

This radio-selected BL Lac \citep{griffith95, landt01} showed a featureless 
optical spectrum when observed in the DXRBS. Our VLT spectrum is also 
featureless (NaI$\lambda$5891 is marginally detected with EW = 0.4 \AA). The
minimum detectable equivalent width is \EWmin$=0.46$ \AA, which gives 
a lower limit $z>0.45$.

\paragraph{RBS 1752} 	      

This X-ray-selected BLL \citep{voges99} had a tentative redshift $z=0.449$
proposed in the Sedentary Survey \citep{giommi05, piranomonte07}, based on 
the possible detection of the host galaxy spectral features in an ESO 3.6 m 
optical spectrum. The high S/N obtained using VLT allowed us to detect 
some weak host galaxy lines (CaII $\lambda\lambda$3934,3968, G band 
$\lambda$4305, and MgI $\lambda$5175) at $z=0.449$. However, the G band is 
possibly contaminated by the [OI] atmospheric line at 6300 \AA, and the 
MgI line is very close to the telluric O$_2$ A band. therefore, while the 
absorption features reported by \citet{piranomonte07} are confirmed, the 
redshift remains tentative because of the lack of other firm absorption 
features.
The fit to the spectrum with a power-law plus elliptical galaxy model gives 
M$_{\rm R}^{\rm host}=-23.3$, in good agreement with the expected distribution 
M$_{\rm R}^{\rm host}=-22.9\pm0.5$.

\paragraph{RBS 1915} 	      

This X-ray selected object \citep{voges99} was classified as a BLL by 
\citet{schwope00}. The optical spectrum reported by \citet{chavushyan00} was 
featureless. Our VLT spectrum shows faint absorption lines from the BLL host 
galaxy (CaII $\lambda\lambda$3934,3968, G band $\lambda$4305, and MgI 
$\lambda$5175) at $z=0.243$. We performed a fit to the spectrum using a 
power-law plus elliptical galaxy model, obtaining M$_R^{host}$=-22.4, consistent
with the expected distribution (M$_R^{host}$=-22.9$\pm$0.5).

\paragraph{TXS 2346+052}	 

This radio selected source \citep{large81} was classified as a BL Lac because 
of its flat radio spectrum \citep{gorshkov00} and the featureless optical
spectrum \citep{chavushyan00}. Our VLT spectrum shows several emission lines 
(MgII $\lambda$2798, [OII] $\lambda$3727, [NeIII] $\lambda$3868, [OIII] 
$\lambda\lambda$4959,5007) at $z=0.419$. The observed EW of the MgII 
and [OIII] lines (exceeding 5 \AA), rules out a BL Lac classification and 
suggests a  FSRQ nature for this source. The EW ratio between the [OII] and 
[OIII] lines are untypical for an AGN, possibly indicating an ongoing star 
formation \citep[as seen in PKS 2005-489 by ][]{bressan06}.
A measurement of the equivalent width of the H$\alpha$+[NII] line system 
(which is out of the observed spectral range) could help to clarify this issue.

\paragraph{1RXS J235730.1$-$171801} 	      

This X-ray selected object \citep{voges99} was classified as a BLL by 
\citet{schwope00}. Our previous VLT observations (S06)
gave a limit $z>$ 0.85. The spectrum presented here has a slightly lower S/N
(110, to be compared with 150 of the earlier observation, because 
of the different seeing conditions between the observations), that gives 
\EWmin$=0.22$ \AA~ and $z>0.60$. CaII$\lambda$3934 and NaI$\lambda$5891
absorptions from the Galaxy ISM are marginally detected (they were also 
detected in the S06 spectrum, along with several DIBs).
No significant flux variations were detected between two different observation 
periods.
\section{Summary and conclusions}

Of 15 observed objects, we confirm the BL Lac classification for 11 
sources, and the detection of a sub-DLA system in PKS 0823$-223$ 
($z\geq0.911$). PKS 1145$-$676  and TXS2346+052 are reclassified as 
FSRQ ($z=0.210$~ and $z=0.419$ respectively), while PMN J1323$-$3652 is a 
F-type star. For 4 BLLs we are able to give a new determination of the redshift
(PKS 1057$-$79 $z=0.569$; RBS 1752 $z=0.448$; RBS 1915, $z=0.243$).
For the remaining 8 BLLs, we give redshift lower limits based on the 
minimum detectable equivalent width of their featureless spectra.
On the whole, our BL Lac spectroscopy database now contains 45 confirmed BL 
Lacs observed with VLT, with 20 redshifts determinated by detection of faint 
lines, and 25 redshift lower limits.

In those cases where even VLT+FORS observations are inconclusive, 
a further increase in the S/N ratio is required, for example through the use of
adaptive optics, Very Large Telescope Interferometry, the Large Binocular 
Telescope, observations in the near infrared region, where the nucleus-to-host 
ratio is smaller than in the optical range, or, in the future, even the use of 
Extremely Large Telescopes. Alternatively, it would be possible to 
observe these sources when the active nucleus goes into a low state, since the 
decrease in the N/H ratio would make easier to detect the features of the host 
galaxy.

\acknowledgments
Based on observations collected at the European Organisation for Astronomical 
Research in the Southern Hemisphere, Chile. Observing proposal: 
ESO 077.B-0045 (PI: S. Ciprini).
SC acknowledges the funding by the European Community's Human Potential 
Programme under contract HPRN-CT-2002-00321.

\newpage
\onecolumn
\begin{figure}[htbp]
    \centering
  \resizebox{\hsize}{!}{\includegraphics{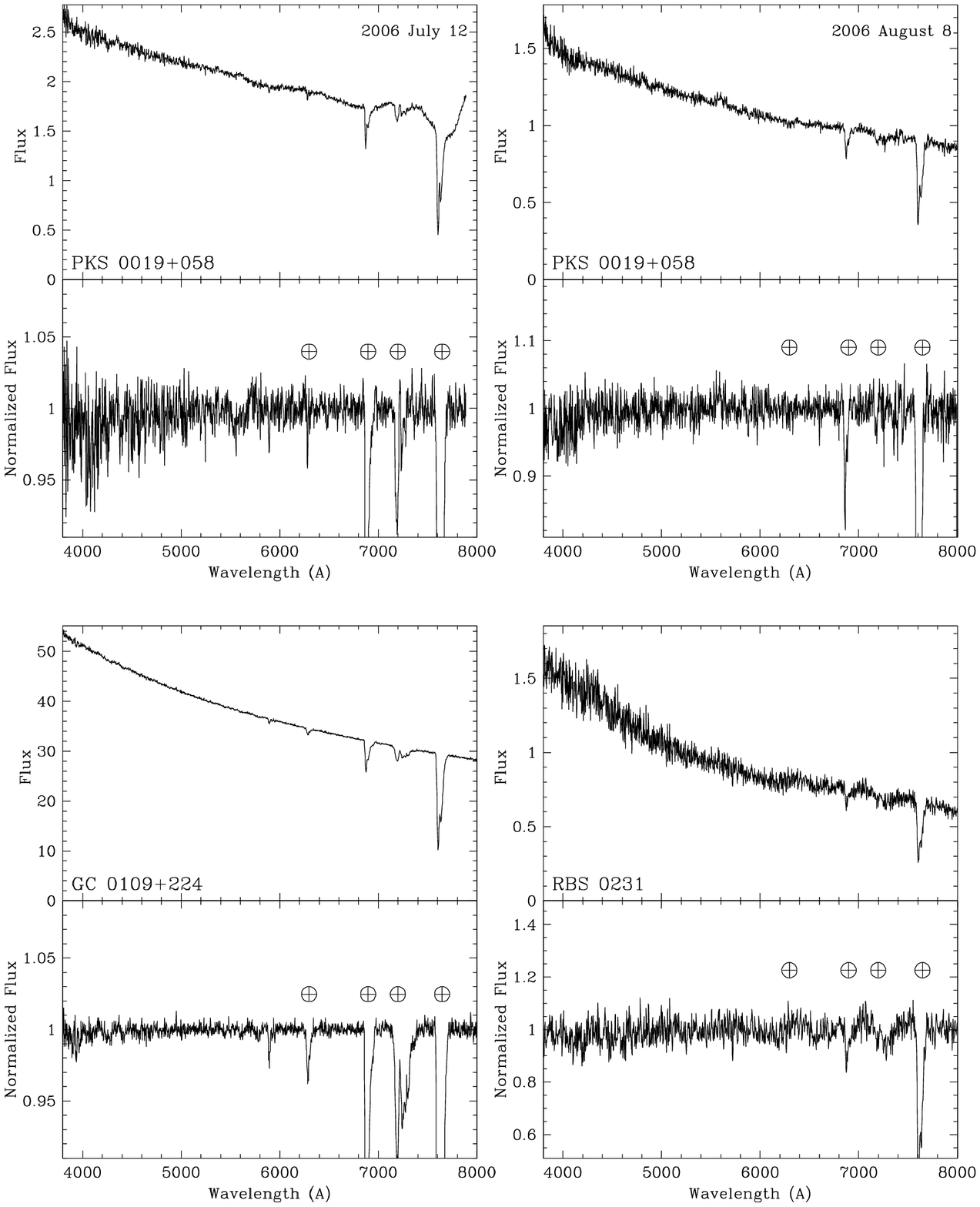}}
  \caption{Spectra of the observed objects. Top panels: flux calibrated 
   dereddened spectra.
   Bottom panels: normalized spectra. 
   Telluric bands are indicated by $\oplus$, spectral lines are marked by the
   line identification, absorption features from atomic species in the 
   interstellar medium of our galaxy are labeled by ISM, diffuse interstellar 
   bands by DIB. The flux \textbf{density} is given in units of $10^{-16}$ erg~ cm$^{-2}$ 
   s$^{-1}$ \AA$^{-1}$}\label{fig:spectra}
\label{fig:spec}
\end{figure}
\addtocounter{figure}{-1}

\begin{figure}[htbp]
    \centering
  \resizebox{\hsize}{!}{\includegraphics{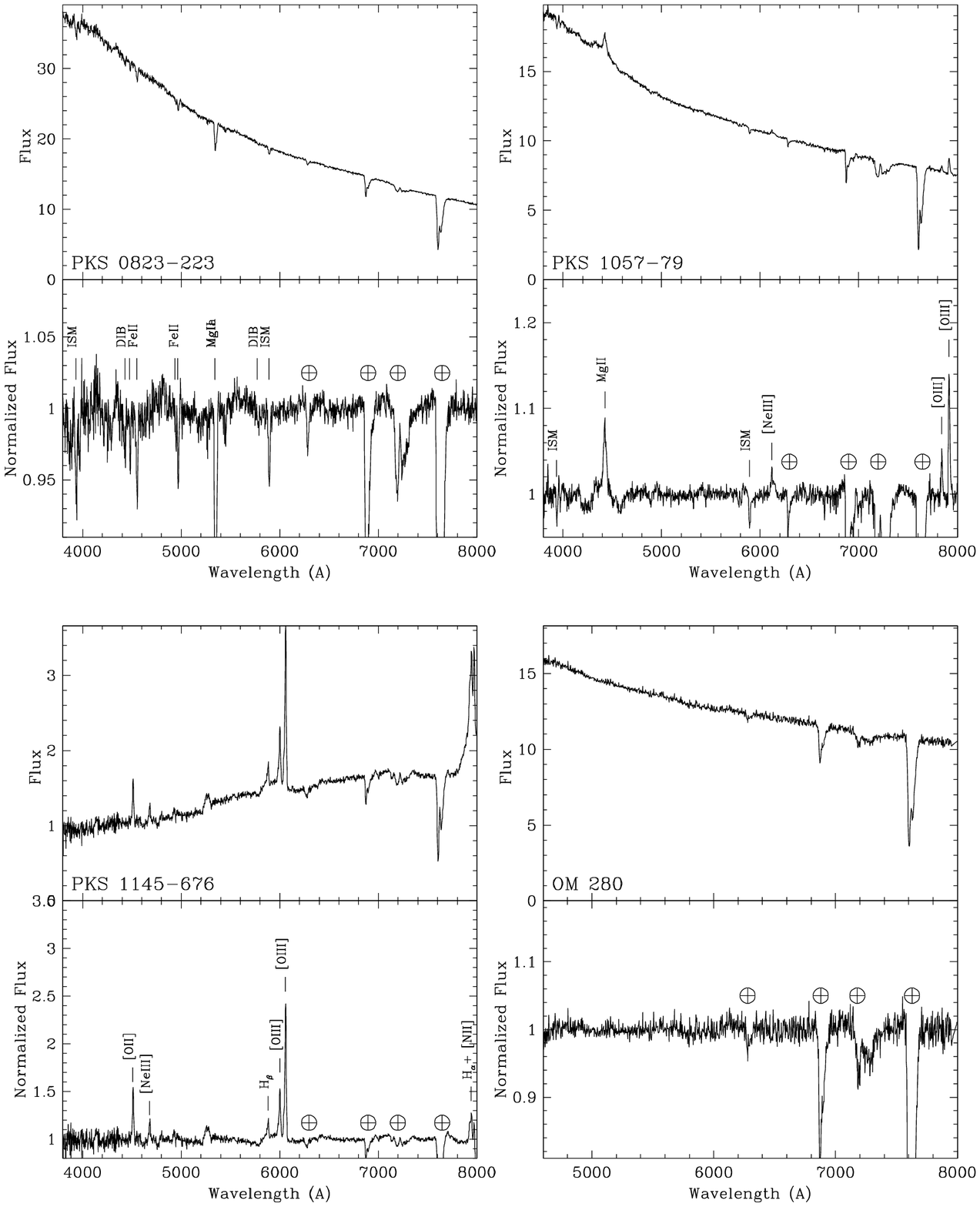}}
  \caption{continued}

\end{figure}
\addtocounter{figure}{-1}

\begin{figure}[htbp]
    \centering
  \resizebox{\hsize}{!}{\includegraphics{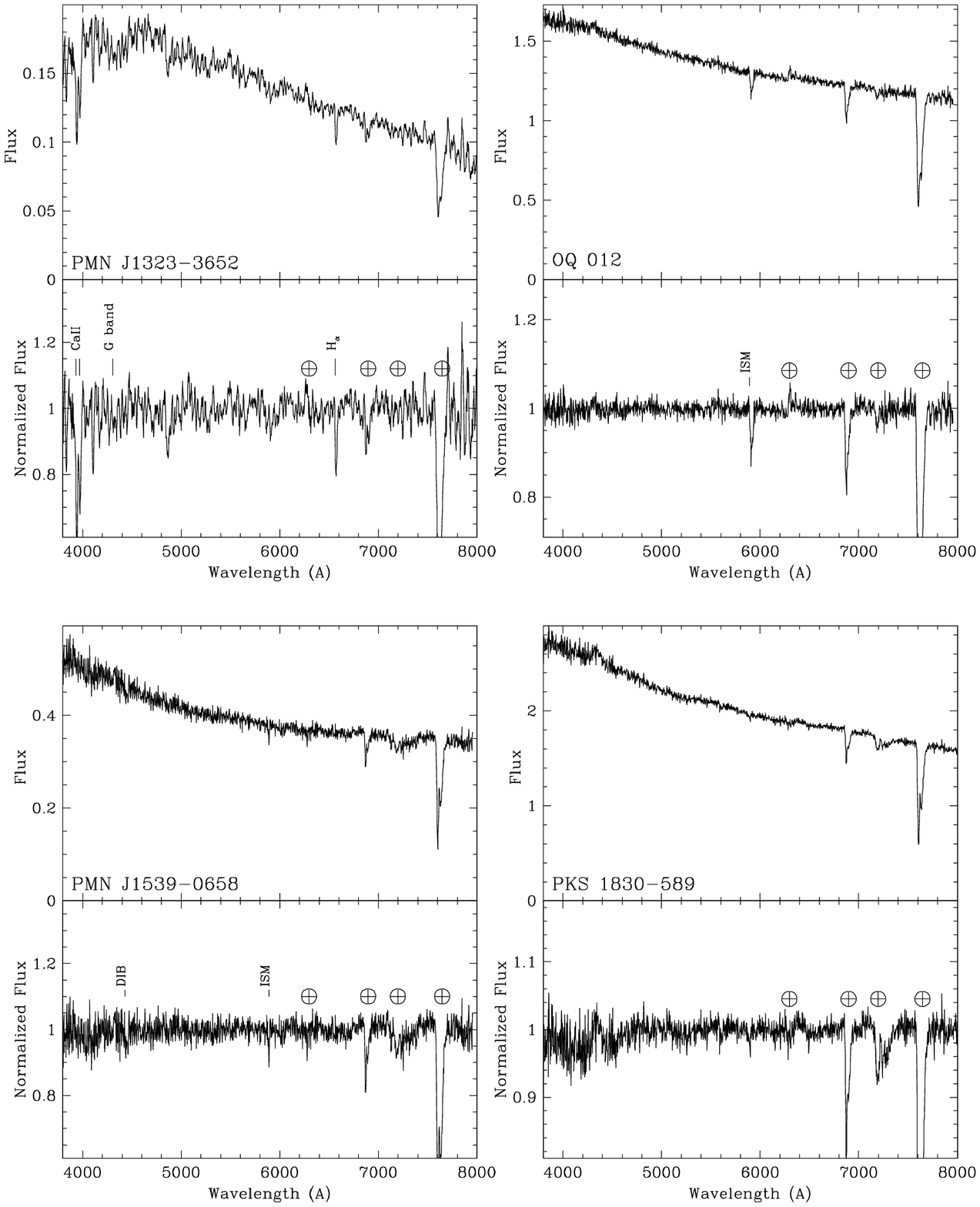}}
  \caption{continued}

\end{figure}
\addtocounter{figure}{-1}

\begin{figure}[htbp]
    \centering
  \resizebox{\hsize}{!}{\includegraphics{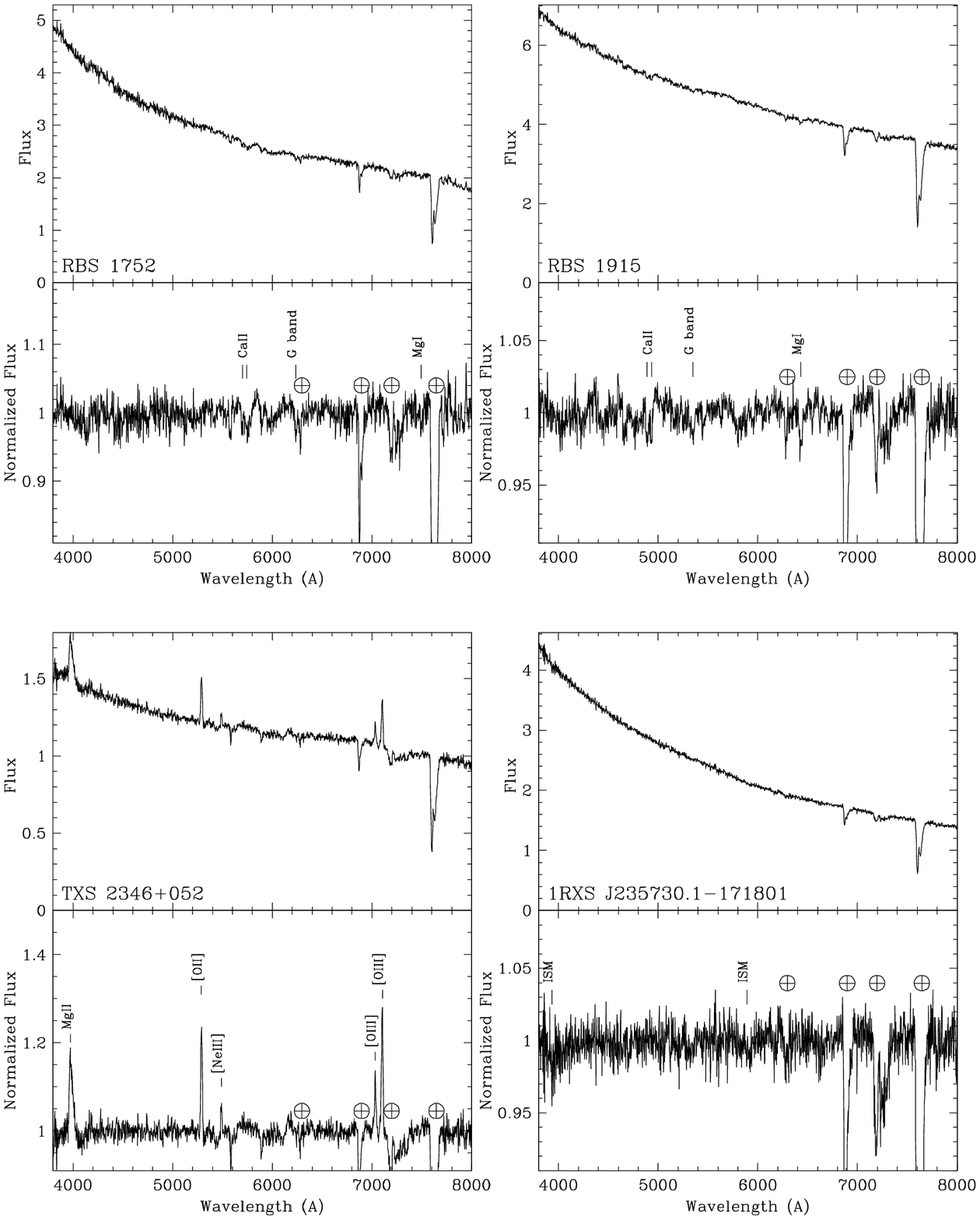}}
  \caption{continued}

\end{figure}

\begin{figure}[htbp]
     \centering
   \resizebox{\hsize}{!}{\includegraphics{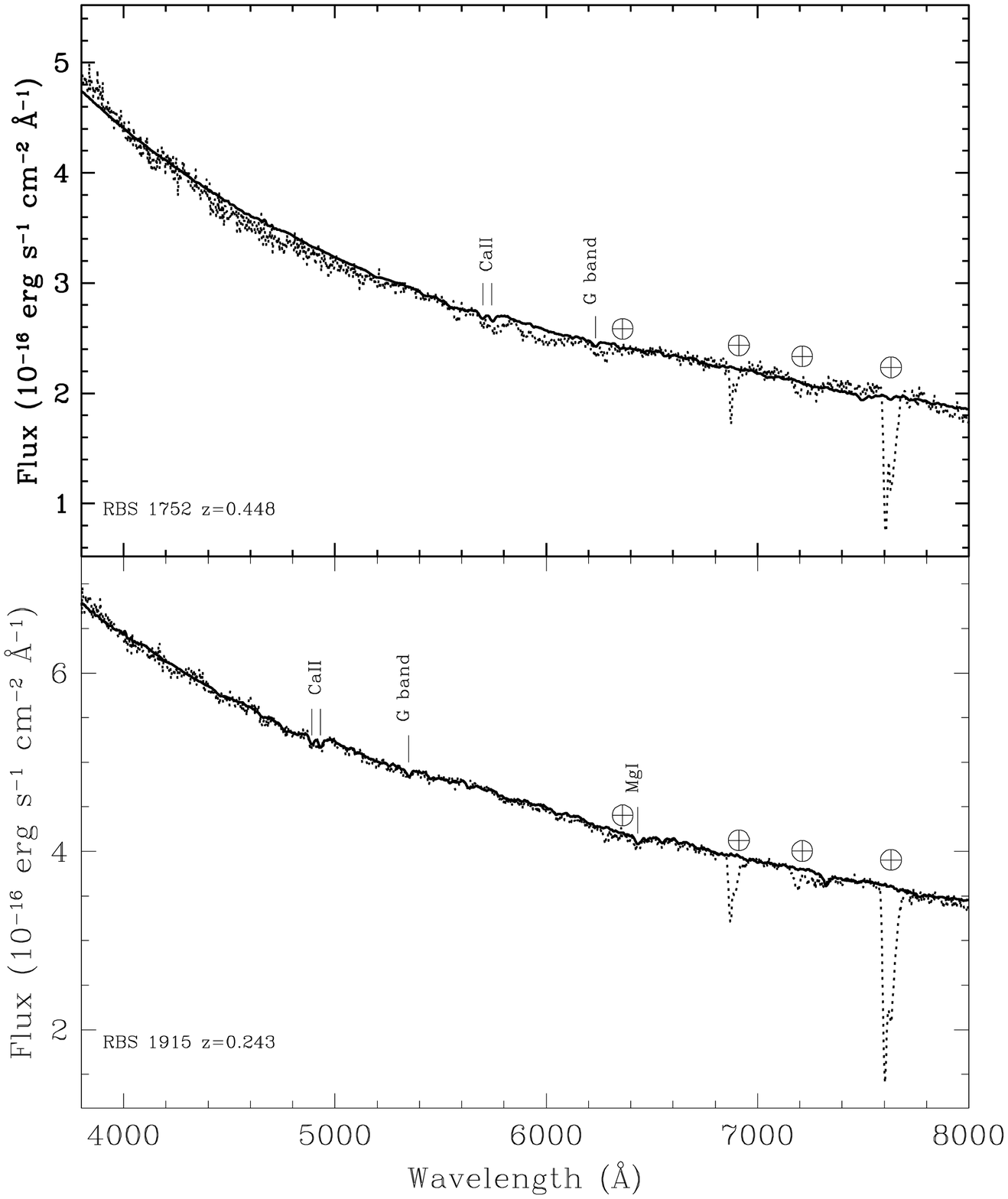}}
   \caption{Spectral decomposition for objects RBS 1752 and RBS 1915. 
   Solid line shows the fitted spectrum, dotted line the observed 
   one.}\label{fig:specdec}

\end{figure}

\newpage

\begin{deluxetable}{lllllllll}
\centering
  \tabletypesize{\footnotesize}
  \tablecaption{Journal of observations\label{tab:journal}}
  \tablehead{
  \colhead{Object name} &  \colhead{IAU name}            &\colhead{RA (J2000)} & \colhead{Dec (J2000)} & \colhead{Date of obs.} & \colhead{Exposure} & \colhead{S/N} & \colhead{\EWmin} & \colhead{$z$} \\
  \colhead{}	        & \colhead{}                     & \colhead{}          & \colhead{}            &  \colhead{}            & \colhead{s}        & \colhead{}    & \colhead{\AA}    & \colhead{}    \\
  \colhead{(1)}	        & \colhead{(2)}                  & \colhead{(3)}       & \colhead{(4)}         &  \colhead{(5)}         & \colhead{(6)}      & \colhead{(7)} & \colhead{(8)}    & \colhead{(9)} \\}
\startdata
PKS 0019+058           &  0019+058 & 00 22 32 & +06 08 04      & Jul 12 	& 2400      & 120   &0.38	    &$>$0.49\\
\nodata                &  ...	   & \nodata  & \nodata        & Aug 08 	& 2400      & 70    &0.40	    &$>$0.64\\  	 
GC 0109+224	       &  0109+224 & 01 12 06 & +22 44 39      & Sep 01 	& 4065      & 380   &0.09	    &$>$0.25\\ 
RBS 0231 	       &  none     & 01 40 41 & $-$07 58 49    & Jul 13 	& 2400      & 20    &1.27	    &$>$0.41\\ 
PKS 0823$-$223	       &  0823-223 & 08 26 02 & $-$22 30 27    & Apr 14 	& 2400      & 220   &0.41	    &$\geqq$0.911\\ 
PKS 1057$-$79	       &  1057-797 & 10 58 43 & $-$80 03 54    & Mar 31 	& 2400      & 90    &0.39	    &0.581\\
PKS 1145$-$676	       &  1145-676 & 11 47 33 & $-$67 53 42    & Apr 14 	& 2400      & 80    &1.84	    &0.210\\
OM 280		       &  1147+245 & 11 50 19 & +24 17 54      & Apr 16 	& 2400      & 100   &0.35	    &$>$0.20\\
PMN J1323$-$3652       &  none     & 13 23 46 & $-$36 53 39    & May 07 	& 2400      & 20    &1.57	    &0\\  
OQ 012		       &  1407+022 & 14 10 04 & +02 03 07      & May 07 	& 2400      & 120   &0.31	    &$>$0.63\\  
PMNJ 1539$-$0658       &  none     & 15 39 10 & $-$06 58 43    & Mar 28, Apr 20 & 4800      & 50    &0.61	    &$>$0.80\\ 
PKS 1830$-$589	       &  1830-589 & 18 34 28 & $-$58 56 36    & Apr 15 	& 2400      & 100   &0.46	    &$>$0.45\\ 
RBS 1752 	       &  none     & 21 31 35 & $-$09 15 23    & May 10 	& 2400      & 90    &0.49	    &0.448\\ 
RBS 1915 	       &  none     & 22 56 13 & $-$33 03 38    & May 05 	& 2400      & 160   &0.35	    &0.243\\ 
TXS 2346+052	       &  2346+052 & 23 49 21 & +05 34 40      & Jul 01 	& 2400      & 80    &0.63	    &0.419\\
1RXS J235730.1$-$171801&  none     & 23 57 30 & $-$17 18 05    & Jul 01 	& 2400      & 110   &0.22	    &$>$0.63\\
\enddata
\tablecomments{Description of columns: 
(1) Object name; 
(2) IAU J1950 code-name;
(3) Right Ascension (J2000); 
(4) Declination (J2000); 
(5) Date of observations, year 2006; 
(6) Exposure time; 
(7) Signal to Noise;
(8) Minimum detectable EW, calculated following \citet{sbarufatti06_vlt2};
(9) Redshifts measured from spectral features and redshift lower limits.
}
\end{deluxetable}

\begin{deluxetable}{lllllll}
\centering
\tabletypesize{\footnotesize}
\tablecaption{Spectral fits\label{tab:fits}}
\tablehead{
  \colhead{Object name}&\colhead{$\alpha$} &\colhead{M$_{\rm R}^{\rm host}$} &\colhead{Class} &\colhead{$R$}   &\colhead{E(B-V)}&\\
  \colhead{(1)}        &\colhead{(2)}      &\colhead{(3)}                  &\colhead{(4)}   &\colhead{(5)}   &\colhead{(6)}   &\colhead{(7)}  \\}
\startdata                      	
PKS 0019+058           & -0.65   &       & L & 17.7 & 0.023 &N96,G00\\
PKS 0019+058           & -0.76   &       & L & 18.4 & 0.023 &N96,G00\\ 
GC 0109+224	       & -0.82   &       & L & 14.6 & 0.049 &N96,W07\\	 
RBS 0231 	       & -1.00   &       & L & 18.6 & 0.109 &G05\\ 
PKS 0823$-$223	       & -0.47   &       & H & 15.4 & 0.472 &D90,W92\\
PKS 1057$-$79	       & -0.73   &       & L & 15.9 & 0.306 &F04,E04\\
OM 280		       & -0.79   &       & L & 15.7 & 0.027 &D01,G95\\
OQ 012		       & -0.29   &       & L & 18.1 & 0.108 &N06\\  
PMNJ 1539$-$0658       & -0.50   &       & L & 19.5 & 0.156 &L01\\  
PKS 1830$-$589	       & -0.65   &       & L & 17.7 & 0.095 &L01\\ 
RBS 1752 	       & -1.45   & -23.3 & L & 17.5 & 0.037 &G05\\ 
RBS 1915 	       & -1.09   & -22.4 & L & 16.8 & 0.018 &B00\\ 
TXS 2346+052	       & -0.06   &       & L\tablenotemark{1} & 18.3 & 0.187 &V05\\ 
1RXS J235730.1$-$171801& -1.40   &       & H & 17.7 & 0.070 &G05\\
\enddata
\tablecomments{
(1) Object name;
(2) Spectral index of the continuum, $\alpha$, defined by 
    F$_{\lambda}\propto\lambda^{-\alpha}$;
(3) absolute $R$ magnitude of the host galaxy;
(4) Class of the object (H: High energy peaked BL Lac, L: Low energy peaked BL Lac);
(5) apparent $R$ magnitude of the object, extracted within a 6''x2'' aperture;
(6) Galactic extinction in the direction of the object, from \citet{schlegel98}.
(7) References for HBL/LBL classification: D90: \citet{dellaceca90}, W92: \citet{white92}, G95: \citet{ghosh95}, N96: \citet{nass96}, B00: \citet{bauer00}, G00: \citet{gorshkov00}, D01: \citet{donato01}, L01: \citet{landt01}, E04: \citet{edwards04}, F04:\citet{flesch04}, G05: \citet{giommi05cat}, V05: \citet{vollmer05}, N06: \citet{nieppola06},  W07: \citet{wu07}.
}
\tablenotetext{1}{Source lacking an X-ray detection in literature. The LBL classification is tentative.}
\end{deluxetable}

\begin{deluxetable}{lllclllll}
\tabletypesize{\footnotesize}
\tablecaption{Spectral lines.\label{tab:lines}}
\tablehead{
\colhead{Object name}   & \colhead{Object Class}& \colhead{$z$} & \colhead{Line ID}	& \colhead{Wavelength} 	& \colhead{$z_{\rm line}$} & \colhead{Type}	&\colhead{FWHM}        & \colhead{EW} \\
\colhead{}              & \colhead{}    	& \colhead{}    & \colhead{}		& \colhead{\AA}      	& \colhead{}		  & \colhead{}    	&\colhead{km s$^{-1}$} & \colhead{\AA} \\
\colhead{(1)}           & \colhead{(2)} 	& \colhead{(3)} & \colhead{(4)}		& \colhead{(5)}       	& \colhead{(6)}		  & \colhead{(7)}    	&\colhead{(8)}         & \colhead{(9)}\\
}
\startdata
PKS 0823$-$223	& sub-DLA/BLL		&$\geqq$0.911    &			& 	&	 &   &	      	&	\\
	        &			&		 & Galactic CaII K 	& 3935  & 0	 & i & 1700    	& 1.85  \\
	        &			&		 & Galactic CaII H 	& 3970  & 0	 & i & 1500    	& 1.00  \\
	        &			&		 & FeII 	   	& 4481  & 0.911  & a & 800     	& 0.47  \\
	        &			&		 & FeII 	   	& 4553  & 0.911  & a & 1300    	& 1.33  \\
	        &			&		 & FeII 	   	& 4948  & 0.914  & a & 600     	& 0.30  \\
	        &			&		 & FeII 	   	& 4970  & 0.912  & a & 800     	& 0.85  \\
	        &			&		 & MgII 	   	& 5349  & 0.911  & a & 1300    	& 4.12  \\
	        &			&		 & MgI  	   	& 5452  & 0.911  & a & 600     	& 0.44  \\
	        &			&		 & Galactic NaI    	& 5893  & 0	 & i & 900     	& 1.08  \\
PKS 1057$-$79	& BLL			& 0.581	 	 & 		   	& 	&	 &   &  	&	\\
                &			&		 & MgII   	        & 4423  & 0.581  & e & 3400  	& -4.24 \\
		&			&		 & NeIII	   	& 6119  & 0.582  & e & 300   	& -0.25 \\
		&			&		 & [OIII]	   	& 7842  & 0.581  & e & 500   	& -1.25 \\
		&			&		 & [OIII]	   	& 7917  & 0.581  & e & 500   	& -3.58 \\
PKS 1145$-$676  & QSO			& 0.210		 & 		   	& 	&	 &   &  	&	\\
		&			&		 & [OII]		& 4512	& 0.210	 & e & 1000     & 9.25  \\
		&			&		 & [NeIII]		& 4680	& 0.210	 & e & 1200     & 3.99  \\
		&			&		 & H$\beta$		& 5880	& 0.210	 & e & 900    	& 5.52  \\
		&			&		 & [OIII]		& 6001	& 0.210	 & e & 1000   	& 9.14  \\
		&			&		 & [OIII]		& 6059	& 0.210	 & e & 800    	& 24.66 \\
		&			&		 & H$\alpha$		& 7944	& 0.210	 & e & \nodata	& \nodata \\
		&			&		 & NII			& 7970	& 0.210	 & e & \nodata	& \nodata \\
RBS 1752	& BLL			& 0.448	 	 & 		   	& 	&	 &   &  	&	\\
		&			&		 & CaII 	   	& 5693  & 0.447  & g & 1000     & 0.5   \\
		&			&		 & CaII 	   	& 5749  & 0.449  & g & 1900     & 0.9   \\
		&			&		 & G band	   	& 6237  & 0.449  & g & 1200     & 1.0   \\
		&			&		 & MgI  	   	& 7493  & 0.448  & g & 600     	& 0.4   \\
RBS 1915	& BLL			& 0.243	  	 & 		   	& 	&	 &   &  	&	\\
		&			&		 & CaII 	   	& 4890  & 0.243  & g & 1700     & 0.75  \\
		&			&		 & CaII 	   	& 4932  & 0.243  & g & 1100     & 0.46  \\
		&			&		 & G band	   	& 5351  & 0.243  & g & 2600     & 0.88  \\
		&			&		 & MgI  	   	& 6429  & 0.243  & g & 1900     & 1.96  \\
TXS 2346+052    & FSRQ			& 0.419	 	 & 		   	& 	&	 &   &  	&	\\
		&			&		 & MgII 	   	& 3973  & 0.420  & e & 3600     & -7.0  \\
		&			&		 & [OII]	   	& 5290  & 0.419  & e & 1000     & -5.0  \\
        	&			&		 & [NeV]	   	& 5488  & 0.419  & e & 800    	& -1.2  \\
        	&			&		 & [OIII]	   	& 7033  & 0.418  & e & 700      & -2.1  \\
        	&			&		 & [OIII]	   	& 7103  & 0.419  & e & 700      & -5.3  \\
\enddata			
					      	       
\tablecomments{Description of columns: 
(1) Object name;     
(2) Object class; 
(3) average redshift;
(4) line identification; 
(5) observed wavelength of line center; 
(6) redshift of the line; 
(7) type of the line (\textbf{e}: emission line, 
\textbf{g}: absorption line from the host galaxy,
\textbf{a}: absorption line from intervening systems), 
\textbf{i}: absorption line from our Galaxy ISM); 
(8) FWHM of the line; 
(9) EW of the line; 
}
\end{deluxetable}
\end{document}